\documentclass[journal]{IEEEtran}

\usepackage{cite}
\usepackage{hyperref} 
\hypersetup{hidelinks}
\usepackage{graphicx}%
\usepackage{multirow}%
\usepackage{amsmath,amssymb,amsfonts}%
\usepackage{amsthm}%
\usepackage{mathrsfs}%
\usepackage{xcolor}%
\usepackage{textcomp}%
\usepackage{manyfoot}%
\usepackage{booktabs}%
\usepackage{algorithm}%
\usepackage{algorithmicx}%
\usepackage{algpseudocode}%
\usepackage{listings}%

\usepackage{balance}
\usepackage{tabularx,booktabs}
\usepackage{amsmath}
\usepackage{array} 
\usepackage{booktabs}

\usepackage{multirow, graphics}

\begin{document}

\title{Crafting Physical Adversarial Examples by Combining Differentiable and Physically Based Renders
}

\author{
   Yuqiu~Liu,
        Huanqian~Yan,
        Xiaopei~Zhu,
        Xiaolin~Hu,~\IEEEmembership{Senior~Member,~IEEE,}
        Liang~Tang,
        Hang~Su,~\IEEEmembership{Member,~IEEE,}
        and~Chen~Lv,~\IEEEmembership{Senior~Member,~IEEE}
\thanks{Yuqiu Liu and Liang Tang are with the School of Technology, Beijing Forestry University, Beijing 100083, China (e-mail: yuqiu\_liu@sfu.ca; happyliang@bjfu.edu.cn).

Huanqian Yan, Xiaopei Zhu, Hang Su are with the Department of Computer Science and Technology, Tsinghua University, Beijing 100084, China (e-mail: yanhq@buaa.edu.cn; zxp18@tsinghua.org.cn; suhangss@mail.tsinghua.edu.cn).

Xiaolin Hu is with the Department of Computer Science and Technology, BNRist, IDG/McGovern Institute for Brain Research, THBI, Tsinghua University, Beijing 100084, China, and also with the Chinese Institute for Brain Research (CIBR), Beijing 100010, China (e-mail: xlhu@tsinghua.edu.cn).

Chen Lv is with the School of Mechanical and Aerospace Engineering, Nanyang Technological University, Singapore 119077, Singapore (e-mail: lyuchen@ntu.edu.sg).

(corresponding authors: Liang Tang and Hang Su.)

{\color{blue}DOI: 10.1109/JAS.2025.125438}
}

}

\markboth{THIS PAPER HAS BEEN ACCEPTED BY \textbf{IEEE/CAA JOURNAL OF AUTOMATICA SINICA}.}%
{Shell \MakeLowercase{\textit{et al.}}: Bare Demo of IEEEtran.cls
for Journals}

\maketitle

\begin{abstract}
Recently we have witnessed progress in hiding road vehicles against object detectors through adversarial camouflage in the digital world. The extension of this technique to the physical world is crucial for testing the robustness of autonomous driving systems. However, existing methods do not show good performances when applied to the physical world. 
This is partly due to insufficient photorealism in training examples, and lack of proper physical realization methods for camouflage. To generate a robust adversarial camouflage suitable for real vehicles, we propose a novel method called \textit{PAV-Camou}. 
We propose to adjust the mapping from the coordinates in the 2D map to those of corresponding 3D model. This process is critical for mitigating texture distortion and ensuring the camouflage's effectiveness when applied in the real world. 
Then we combine two renderers with different characteristics to obtain adversarial examples that are photorealistic that closely mimic real-world lighting and texture properties. The method ensures that the generated textures remain effective under diverse environmental conditions.
Our adversarial camouflage can be optimized and printed in the form of 2D patterns, allowing for direct application on real vehicles.
Extensive experiments demonstrated that our proposed method achieved good performance in both the digital world and the physical world.
\end{abstract}


\IEEEpeerreviewmaketitle

\section{Introduction}

\IEEEPARstart{A}{dversarial} attacks represent a critical vulnerability in Deep Neural Networks (DNNs), capable of manipulating DNNs to produce incorrect results through carefully designed perturbations in inputs, and the perturbed inputs are called {\it adversarial examples}. 

Initially identified in the digital world, adversarial attacks have been extended to the physical world, underscoring a significant security concern \cite{goodfellow2014explaining, carlini2017towards, madry2017towards, dong2018boosting, zhou2022modeling}. 
Adversarial examples can deceive DNNs in both visible light fields \cite{huang2020universal, xu2020adversarial, hu2022adversarial, thys2019fooling, wang4545326naturalistic, sun2023differential} and infrared fields \cite{zhu2022infrared, zhu2021fooling, wei2023hotcold}.

\begin{figure}[t]

  \setlength{\abovecaptionskip}{0 pt}  
  \setlength{\belowcaptionskip}{0.0cm} 
  \begin{center}
  \includegraphics[width=1.0\linewidth]{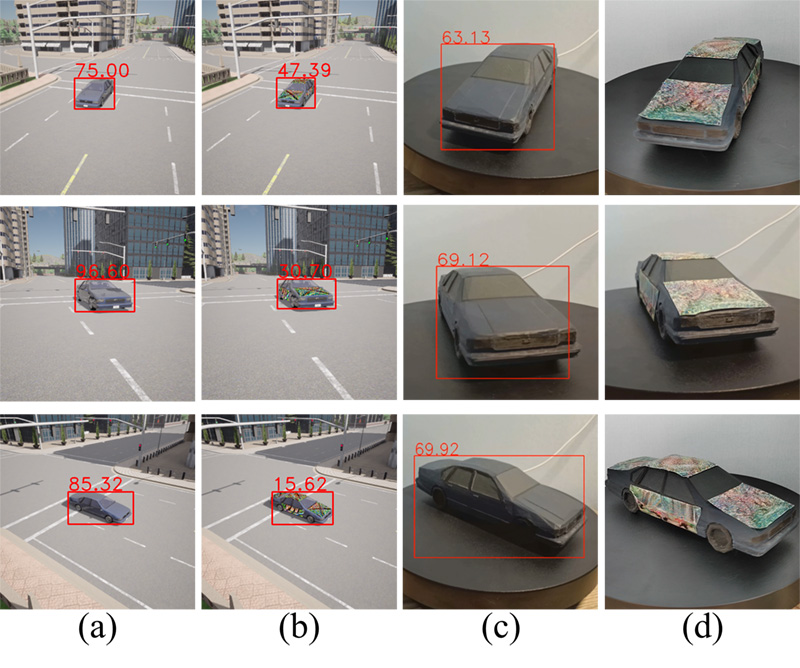}
  \end{center}
  \caption{Detection results of the normal car and the camouflaged car with different viewing angles. The red score (\%) is the confidence of the detector Faster R-CNN. Usually attacks are regarded successful if the score is lower than 50\%. (a) The normal car was detected successfully. (b) The camouflaged car significantly decreased the detection confidence in the digital world. (c) The normal car model in the physical world was detected. (d) The physical car model with our adversarial textures attached on the surface was not detected. More examples are shown in Fig. \ref{fig:digital} and Fig. \ref{fig:physical}.}
\label{fig:onecol}

\end{figure}

2D image-based attacks generate 2D patterns or optimize positions of fixed patches that can be applied to the target object to deceive DNN detectors \cite{thys2019fooling, wei2024revisiting}. Usually they could only deceive DNNs in a limited range of views. 
One exception is a recent work \cite{hu2022adversarial} where certain basic 2D pattern was tiled on a large piece of cloth then the target object was covered with this cloth. With carefully designed 2D pattern, a multi-view adversarial attack was successfully achieved: persons wearing the cloth could be hardly detected by DNN-based person detectors. However, its effect is limited since the basic 2D pattern is required to have adversarial attack effect at any part of the object viewed from any viewing angle. It poses difficulty to hide objects from detectors which have quite different appearances at different viewing angles (e.g., cars). 

The attacks based on 3D models aim to generate adversarial textures that can be attached to the surfaces of target objects \cite{wu2020physical, hu2023physically, wang2021dual,wang2022fca, zhang2023boosting, duan2021learning, suryanto2022dta, suryanto2023active}. 
The process allows the camouflaged objects to get a score lower than detection threshold from multiple angles through attacks. 
Unfortunately, the insufficiently photorealistic rendering results have constrained the performance of 3D adversarial camouflage in the physical world. While existing methods \cite{wang2021dual,wang2022fca} use differentiable renderers to support the optimization of their adversarial camouflages, these renderers often sacrifice some complex rendering processes for differentiability. The challenge lies in how to ensure photorealistic rendering results of vehicles when generating adversarial textures through gradient backpropagation.
In this work, we aim to achieve realizable and robust camouflage in the multi-view real-world setting.

The primary challenge for 3D camouflage attacks lies in the rendering of 3D objects. Physically Based Rendering (PBR), a widely adopted rendering method, can produce images that closely resemble photos of the real world. Considering that PBR lacks backpropagation support, Wu \textit{et al.} \cite{wu2020physical} attempt to generate adversarial camouflages using a discrete searching method on PBR, but the large color space makes it challenging to find the optimal solution, leading to unsatisfactory attack performance. 
Other methods \cite{hu2023physically, wang2021dual,wang2022fca, zhang2023boosting, duan2021learning, suryanto2022dta, suryanto2023active} based on 3D modeling employ differentiable renderers (DR) for rendering. The major merit of DR is that it supports backpropagation to optimize the textures of 3D models, but to achieve guideability, DR has to abandon some advanced rendering techniques and simplify or approximate the calculation of the rendering process. This simplification makes it difficult for DR to achieve the rendering realism of PBR. 

Physical realizability is also an important factor in designing physical attack methods. Most methods \cite{wang2021dual,wang2022fca, suryanto2022dta, zhang2023boosting, duan2021learning} generate camouflages that are attached to 3D models of target objects. These textures can only be fully reproduced in the physical world through techniques like 3D printing or 3D spraying. Usually 3D printing cannot be applied to existing objects in the physical world, e.g., a real car. Furthermore, due to the limitations of 3D spraying technology, effectively reproducing textures on the surfaces of real objects is challenging. Hu \textit{et al.} \cite{hu2023physically} use a vertical projection mapping from 3D model to 2D patterns to ensure the feasibility of physical implementation. However, this method is only suitable for relatively flat surfaces. See Section \ref{adjust} for detailed analysis.

In this paper, we present an attack method to generate a physical adversarial vehicle camouflage, which is called {\it PAV-Camou}.
We combine DR and PBR to generate images that not only support optimization but also closely resemble real vehicles. 
To facilitate physical realization on existing objects, before rendering, we propose to introduce an additional step. In this step, we align the mapping from the coordinates in the 2D map to those of corresponding 3D model to reduce texture deformation. Experiments showed that the proposed method was effective in hiding vehicles against object detectors in both the digital world and the physical world.

\begin{figure*}[t]
   \begin{center}
   \includegraphics[width=1.0\linewidth]{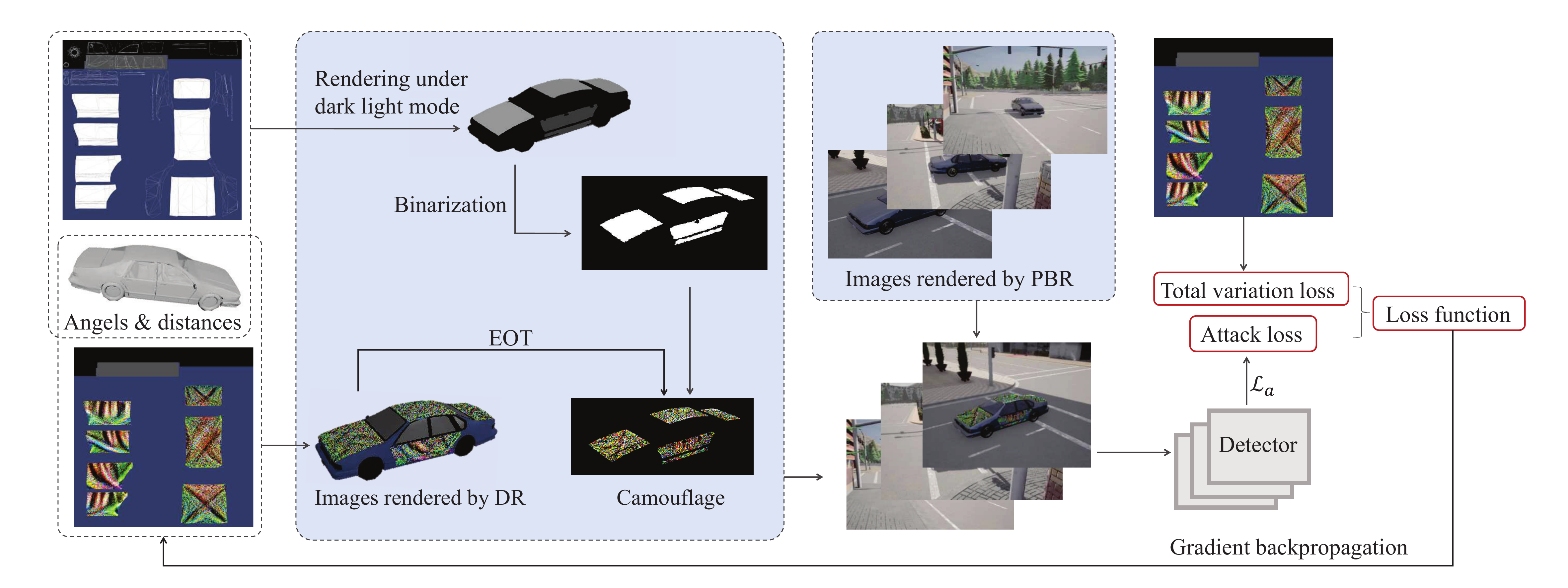}
   \end{center}
      \caption{Overview of the optimization pipeline. DR renders the vehicle with adversarial textures, and PBR renders the vehicle with original textures. Their results are combined using masks, which are obtained by rendering the vehicle under dark light mode and binarizing the results. The textured parts of DR results are taken for gradient backpropagation, and the non-textured parts of PBR results are taken for a photorealistic appearance. In this way, adversarial textures can be tailored for attacking real vehicle detectors.
   }
   \label{fig:Overview}
   \end{figure*}

\section{Related work}\label{sec2}

In this section, we first provide an overview of recent works on physical adversarial attacks, including typical camouflage-based 3D adversarial attack methods, and then introduce some renderers.

\subsection{Physical Adversarial Attacks}
Initially, the majority of adversarial research efforts concentrated on the digital domain, investigating vulnerabilities and developing attack methodologies for computer vision models \cite{szegedy2013intriguing, moosavi2017universal, moosavi2016deepfool,su2019one}. To realize physical adversarial attacks, some methods proposed enhancement strategies to bridge the gap between the digital and physical worlds \cite{athalye2018synthesizing, eykholt2018robust, li2019adversarial,xu2020adversarial}.

Athalye \textit{et al.} \cite{athalye2018synthesizing} introduced a notable method to generate adversarial examples in the physical world. 
Their method, known as Expectation over Transformation (EoT), introduces disturbances such as brightness, contrast, and other transformations to adversarial noise during optimization.
These operations ensure that the images retain certain adversarial features when printed and placed in the real world. The EoT operation has become a basic technique for numerous studies on physical adversarial attacks \cite{hu2022adversarial, duan2020adversarial,chen2019shapeshifter}. Additionally, in pursuit of improved physical attack performance, Xu \textit{et al.} \cite{xu2020adversarial} proposed to utilize Thin Plate Spine (TPS) to simulate the distortion of the object surfaces. 
However, most of these attacks are only applicable to specific scenarios. 

To achieve more practical physical adversarial attacks, some works explored camouflage-based adversarial attacks in 3D scenes \cite{zhang2019camou, wu2020physical, wang2021dual,wang2022fca, zhang2023boosting, duan2021learning}. 
Camouflage-based attacks aim to generate textures attached to a 3D model, enabling the object to be concealed from multiple perspectives.
Zhang \textit{et al.} \cite{zhang2019camou} trained an agent network to predict detection scores using Physically-Based Rendering (PBR) and optimize vehicle textures based on those predictions. 
However, the agent network did not actually render the models, which creates a gap between the predicted results and the actual rendering outcomes. 
Wu \textit{et al.} \cite{wu2020physical} introduced adversarial camouflage based on genetic algorithms, but traditional search algorithms struggled to find the optimal solution for adversarial camouflage.
In recent years, the emergence of 3D differentiable renderers has facilitated the optimization of 3D camouflage. Wang \textit{et al.} \cite{wang2021dual} and Zhang \textit{et al.} \cite{zhang2023boosting} optimized adversarial patterns using DR. Wang \textit{et al.} \cite{wang2022fca} presented an end-to-end framework for 3D digital camouflage optimization. However, these methods faced challenges in effective physical application. Suryanto \textit{et al.} \cite{suryanto2022dta} and Duan \textit{et al.} \cite{duan2021learning} used 3D printing for application, but due to the lack of proper 2D mapping, these textures attached to 3D models could not be applied to real vehicles.

To solve the aforementioned problems, we propose a mapping method from 2D textures to 3D models and combine PBR and DR for vehicle rendering. It allows the 3D camouflage textures to be realized in the physical world while maintaining attack effectiveness.

\subsection{Rendering} 
Rendering is the process of generating an image from a 3D model by means of specialized computer programs known as renderers.
Early renderers aimed to produce more realistic images by simulating the physical phenomena of light reflection in the real world, known as Physically Based Rendering (PBR) \cite{sanders2016introduction}. 
However, due to the intricate nature of light reflection in the physical world, PBR was either non-differentiable or too complex to generate adversarial textures.
In recent years, with the advancement of machine learning, some methods were proposed to achieve differentiable rendering (DR) \cite{kato2018neural, ravi2020accelerating} using an approximating rasterization. 
In order to enable gradient backpropagation, DR sacrifices many physical characteristics of the rendering process and replaces them with simplified computations \cite{meloni2021messing}. 
As a result, while DR allows for gradient backpropagation, the rendered images are still far from resembling objects in the real world.

Before the emergence of DR, physical adversarial attacks on 3D objects were primarily achieved through search methods or by training proxy networks. After the advent of DR, most approaches have been implemented based on DR. There are two common forms of texture representation in DR: one where the texture is directly attached to the model based on faces, and the other where the texture is based on 2D maps and rendered through U-V mapping. To achieve adversarial attacks, these textures need to be optimized. Methods such as DAS \cite{wang2021dual}, FCA \cite{wang2022fca}, and TPA \cite{zhang2023boosting} use face-based texture rendering, while CAC \cite{duan2021learning} uses map-based texture rendering. 

In this work, we adopt map-based texture rendering and redesign the U-V mapping accordingly.

\section{The PAV-Camou attack method}
We start with formulating the adversarial camouflage problem and providing an overview of the proposed method. 
Then we present the process of 2D coordinate adjustments and the combination of PBR and DR. 
Finally, we introduce our loss function and describe the optimization process.
\subsection{Problem Formulation}
The goal of the adversarial attack is to decrease the detection score of the target class. 
To generate textures attached to the 3D model, we perform an adversarial attack on the confidence scores of the rendered images $\mathbf{I}$.
The rendering process is depicted as a function $\mathcal{R}$ that produces a image $\mathbf{I}$: 
\begin{equation}
\mathbf{I}=\mathcal{R}(\Theta, \mathbf{M}, \mathbf{S}),
\label{eq:render}
\end{equation}
where $\Theta$ denotes coordinates $(r, \theta, \varphi)$ of the visible light sensor in a spherical coordinate system; $\mathbf{S}$ denotes the scene in the simulation environment; $\mathbf{M}$ denotes the 3D model of the vehicle, which is composed of the blank model $\mathbf{M}_0$ and the texture $\mathbf{T}$, i.e., $\mathbf{M}=(\mathbf{M}_0,\mathbf{T})$. 

The adversarial texture map $\mathbf{T}_{adv}$ is employed to deceive the victim detector $\mathcal{F}$. Hence, the problem can be described as:
\begin{equation}
\underset{\mathbf{T}_{adv}}{\arg\min}\quad\underset{\Theta,S}{\mathbb{E}}\ [\mathcal{F}(\mathcal{R}(\Theta, (\mathbf{M}_0,\mathbf{T}_{adv}), \mathbf{S}))],
\end{equation}
where $\mathbb{E}$ denotes the expectation of confidences.


\begin{figure*}[h]
   \begin{center}
   \includegraphics[width=0.8\linewidth]{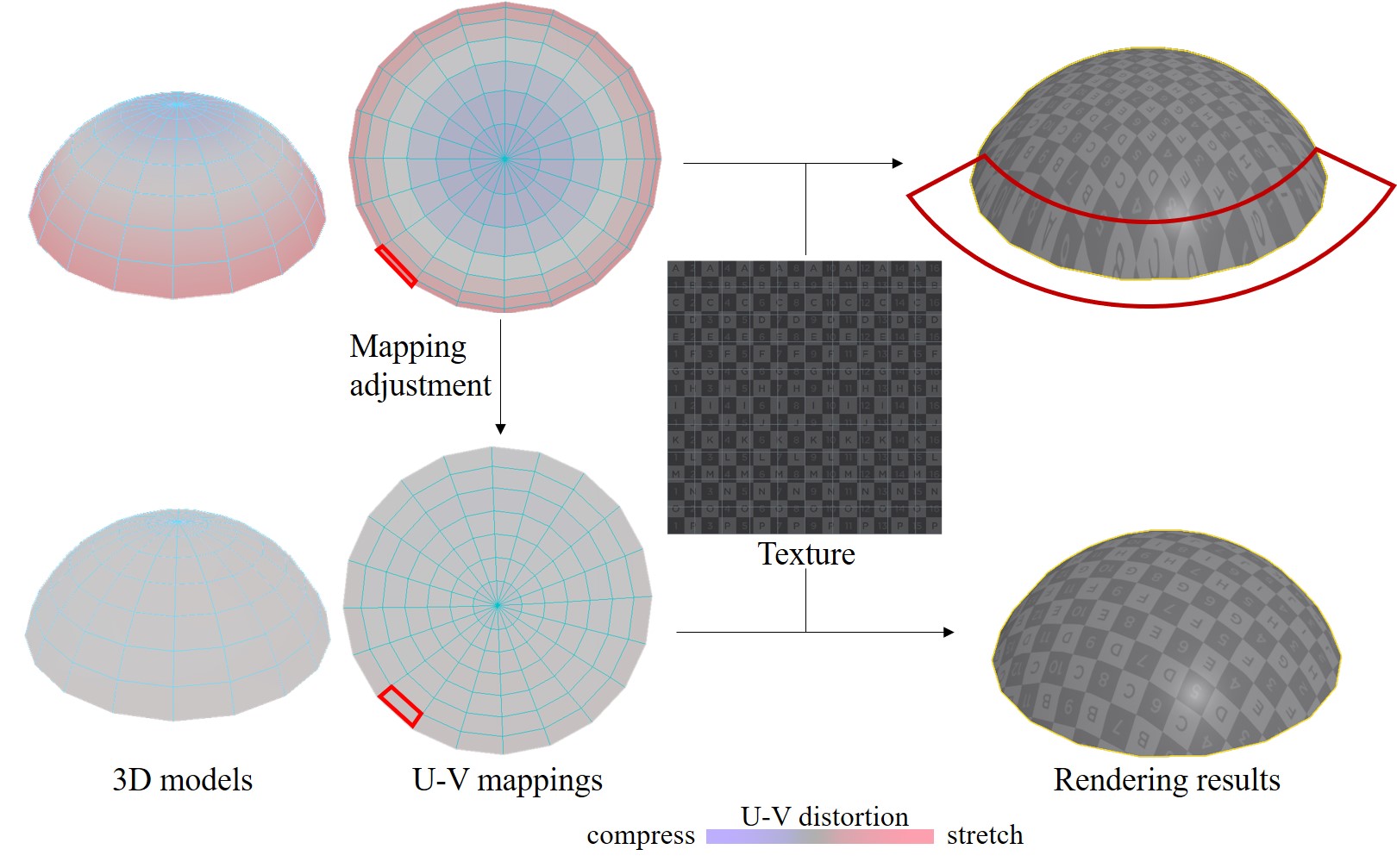}
   \end{center}
      \caption{Different U-V mappings of two same hemispheres and their rendering results. The first column shows two identical 3D models (hemispheres). The second column shows the U-V map, where the two trapezoids circled represent the mapping results of the same quadrilateral before and after 2D adjustments. The last column shows the rendered hemispheres after mapping. As shown in the top row, generating 2D coordinates of mapping by the direct projection can lead to texture distortion, and in the bottom row, by adjusting the coordinates, the distortion is reduced.
   }

   \label{fig:problem}
   \end{figure*}
   
\subsection{Overview of Our Method}

We first adjust the mapping from 2D coordinates to the 3D model, which is also called U-V mapping. 
A corresponding image is then generated based on these coordinates, where textured parts are marked.
Next, the 3D model is rendered using both PBR and DR. 
PBR is used for rendering the environment and non-textured parts of the vehicle, as it supports complex rendering processes and environment setup. DR is used for rendering the textured parts from multiple angles, as it enables gradient backpropagation. 
The combination of PBR and DR reduces the impact of non-textured parts on adversarial texture formation, resulting in more effective adversarial textures on real vehicles.
Finally, several loss functions are employed to optimize the marked parts in the 2D map.
The overall pipeline is shown in Fig. \ref{fig:Overview}.

\subsection{2D Coordinate Adjustments}
\label{adjust}
U-V mapping is the process of projecting 2D coordinates onto a 3D model to create texture mapping. The 3D object is unwrapped and the 2D texture is applied to it. 
However, most 3D models lack a well-defined U-V mapping. To achieve camouflage-based attacks, some methods \cite{wang2022fca,wang2021dual} rely on painting on the triangular mesh faces directly. This process poses challenges to physical implementation. To overcome it, they use an original mapping to apply the adversarial textures, which leads to high texture distortion when the textures are applied to curved surfaces.
If an unreasonable mapping method is used, the faces (triangles or quadrilaterals) in the 3D model will be stretched or compressed after mapping to the 3D model. The stretching and compression are expressed by purple and red colors respectively in the 3D modeling software MAYA \cite{derakhshani2012introducing}. These distortions are called U-V distortions in 3D modeling. As shown in the upper part of Fig. \ref{fig:problem}, an unadjusted U-V map causes stretching of the texture, especially in the part with significant U-V distortions (highlighted by the red curves on the top right figure). In the lower part of Fig. \ref{fig:problem}, after coordinates adjustments (see below), the U-V distortions in the model are reduced, indicated by the gray color. 

We propose an adjustment scheme for the 2D coordinates of the mapping, ensuring minimal deformation of textures when the textures are attached to the existing object. We also create a corresponding initial image with marked textured parts, which is called the U-V patch.

We adjust the 2D coordinates of U-V mapping in MAYA. MAYA can visualize U-V distortions in real-time during the adjustment process, facilitating our determination of the new coordinates. The adjustment process is similar to unfolding a 3D model. In this process, we start from the center point of a surface and adjust the coordinates of its neighbors, ensuring that the U-V distortions of the faces with this point as the vertex are minimized. Similarly, we adjust the coordinates of these neighbors' neighbors, and so on, until the U-V distortions of the entire surface are minimized as much as possible. After that, the color of these faces is close to gray in MAYA, as shown in the middle of the bottom row of Fig. \ref{fig:problem}. Note that due to the inability of the surface to become completely 2D, slight U-V distortions are unavoidable.

\begin{figure}[h]
   \begin{center}
   \includegraphics[width=1.0\linewidth]{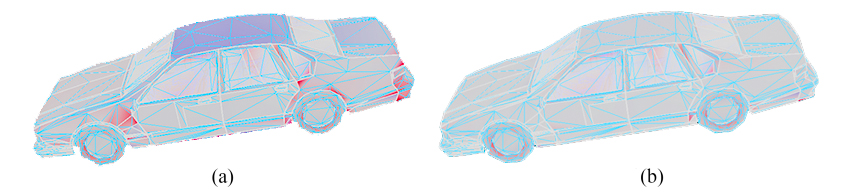}
   \end{center}
      \caption{In the same 3D car model, the U-V distortions before (a) and after (b) adjustments. Purple indicates stretching, red indicates compression, and gray indicates no distortion.}
   \label{fig:map}
   \end{figure}

Please note that this is a preprocessing step for DR. Once the mapping is established, it does not require further adjustment during subsequent rendering and training processes.
The comparison of U-V distortions on our chosen car model before and after adjustments is shown in Fig. \ref{fig:map}.
The whole map is shown on the left of Fig. \ref{fig:Overview}.

\subsection{Combination of Two Renderers}

To ensure the effectiveness of adversarial patterns in the real world, we propose a combination of PBR and DR. The textured parts of the target vehicle are rendered using DR so that they can be optimized through gradient backpropagation. Conversely, non-textured parts do not participate in optimization, so using PBR helps preserve their more photorealistic appearance. 

We render the target 3D model $\mathbf{M}$ from the same viewing angle and distance using both PBR and DR. Then we combine the result $\mathbf{I}_{p}$ rendered with PBR and the result $\mathbf{I}_{d}$ rendered with DR using a mask $\mathbf{P}$ (see Fig. \ref{fig:Overview}). 
The image $\mathbf{I}_{p}$ generated from PBR $\mathcal{R}_p$ can be expressed as
\begin{equation}
\mathbf{I}_{p}=\mathcal{R}_p(\Theta, \mathbf{M}, \mathbf{S}),
\end{equation}
where $\Theta, \mathbf{M}, \mathbf{S}$ are defined in Eqn. (\ref{eq:render}).
During optimizing, we generate the image $\mathbf{I}_d$ with textures $\mathbf{T}_t$ by DR $\mathcal{R}_d$ in real time. 
\begin{equation}
\mathbf{I}_{d}=\mathcal{R}_d(\Theta, (\mathbf{M}_0,\mathbf{T}_t), \mathbf{S}').
\end{equation}

Before combining, textures in $\mathbf{I}_{d}$ are transformed with EoT \cite{athalye2018synthesizing} for robust physical performance. The procedure can be described as $\mathbf{I}_{d}'= \mathcal{E}(\mathbf{I}_{d})$, where $\mathcal{E}$ indicates transformations including brightness, contrast, and so on. We expect that for each $\mathbf{I}_{d}$, there is a corresponding mask $\mathbf{P}$, where the value is 1 for the textured part of $\mathbf{I}_{d}'$ and 0 for the non-textured part. 
With the help of the mask $\mathbf{P}$, we can combine the results of PBR and DR:
\begin{equation}
\mathbf{I}_{o}=\mathbf{I}_{d}'\cdot\mathbf{P}+\mathbf{I}_{p}\cdot(1-\mathbf{P}).
\label{eq:merge}
\end{equation}

\begin{figure}[t]
   \begin{center}
   \includegraphics[width=1.0\linewidth]{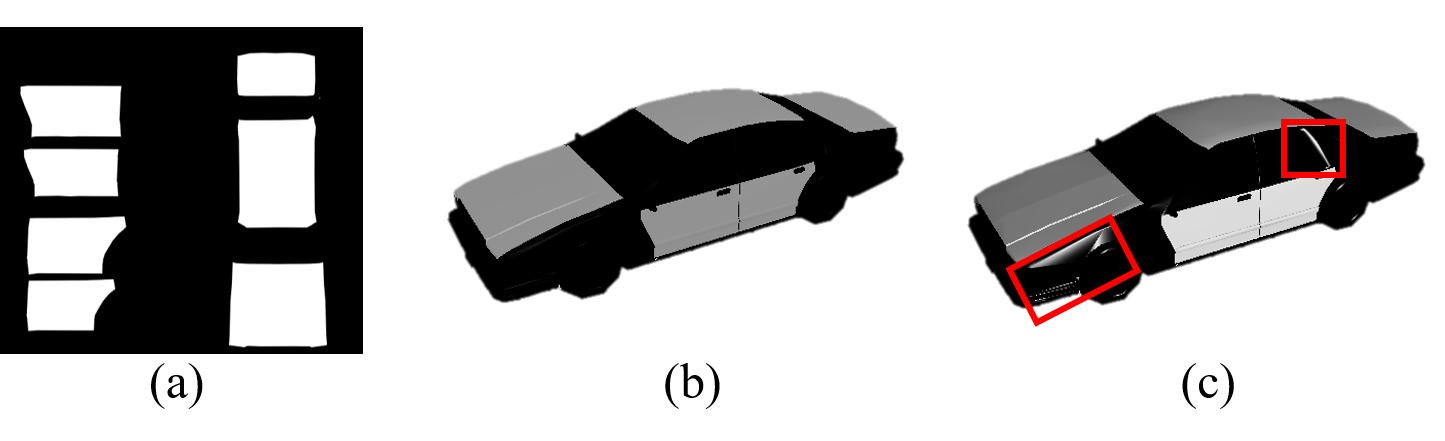}
   \end{center}
      \caption{The U-V map required for mask generation and its DR results under different lighting conditions. (a) The texture map $\mathbf{T}_m$ used for generating the mask. (b) The rendering result $\mathbf{I}_m$ under dark light mode. (c) The rendering result under a strong light mode. In (c),  the red boxes contain certain car body areas that have similar gray values to the car doors.
      }
   \label{fig:dark_render}
   \end{figure}

However, since the rendering process can alter the brightness of the texture, the mask $\mathbf{P}$ cannot be obtained by checking if the DR result $\mathbf{I}_{d}'$ equals the values of the textured part in the U-V map, i.e., $\mathbf{P}$ cannot be derived from $\mathbf{I}_{d}'$. To accurately distinguish between the two parts in $\mathbf{I}_{d}'$, we create another grayscale map $\mathbf{T}_m$ with two values $C_t$ and $C_n$ (see Fig. \ref{fig:dark_render}(a)), rendering it as a texture of the car model in a specific scene $\mathbf{S}'$ to obtain the grayscale image $\mathbf{I}_{m}$ (see Fig. \ref{fig:dark_render}(b)). 

The image $\mathbf{I}_m$ rendered in such settings can be formulated as 
\begin{equation}
\mathbf{I}_m=\mathcal{R}_d(\Theta, \mathbf{M}_m, \mathbf{S}'),
\end{equation}
where $\mathbf{M}_m = (\mathbf{M}_0, \mathbf{T}_m)$.

The gray value of the rendered vehicle is influenced by ambient lighting and its own material. Due to specular reflection, some parts with the same color in the map can exhibit different colors after rendering. Therefore, the two different colors in $\mathbf{T}_m$ are mapped to separate grayscale ranges in $\mathbf{I}_{m}$ instead of two values. 
The grayscale range of the textured parts in the DR results is $(c_t^{min}, c_t^{max})$, and that of the non-textured parts is $(c_n^{min}, c_n^{max})$. Note that the gray scale ranges $(c_t^{min}, c_t^{max})$ and $(c_n^{min}, c_n^{max})$ are obtained by measuring an example in DR results using image editing software instead of calculating based on $C_t$ and $C_n$.
To perform binarization, it is necessary to ensure that $c_t^{min}\geq c_n^{max}$ or $c_t^{max}<c_n^{min}$. In this way, we can find an intermediate value $c^{mid}$ that is not within the two color ranges. 
\begin{equation}
   \label{eq:cmid}
   c^{mid}=\left\{\begin{array}{ll}
   (c_t^{min}+c_n^{max})/2,\quad \text{if}\ c_t^{min}\geq c_n^{max},\\
   (c_t^{max}+c_n^{min})/2,\quad \text{else}.\\
   \end{array}\right.
   \end{equation}

By using the intermediate value $c^{mid}$ as a threshold, the rendered image $\mathbf{I}_{_m i,j}^{c}$ is binarized to obtain the mask $\mathbf{P}=\{p_{_{i,j}}\}$, namely
   \begin{equation}
      \label{eq:P}
   p_{_{i,j}}=\left\{\begin{array}{ll}
   1,\quad \text{if}\ \mathbf{I}_{_m i,j}^{c}\geq c^{mid},\\
   0,\quad \text{else},\\
   \end{array}\right.
   \end{equation}
where $i,j$ denote the $i$-th row, $j$-th column in the image.

In the above rendering process, to ensure either $c_t^{min}\geq c_n^{max}$ or $c_t^{max}<c_n^{min}$, we set the scene $\mathbf{S}'$ to dark light mode. 
The dark light mode refers to the rendering condition without directional lights and point lights. In the dark light mode, parts with the same material have minimal color difference after rendering (see Fig. \ref{fig:dark_render}(b)).
Therefore, parts with different colors can still maintain their color difference after rendering, as they are not affected by reflections that may cause significant color fluctuations. 
The dark light mode ensures that different parts with the same color have minimal color difference after rendering. 
This prevents the two color ranges $(c_t^{min}, c_t^{max})$ and $(c_n^{min}, c_n^{max})$ from intersecting in the rendered results, which facilitates the generation of masks. On the contrary, adopting strong light mode would result in reflections from metallic surfaces causing highlights on car bodies, leading to the intersection of the two color ranges in the rendered results, which means $(c_t^{min}, c_t^{max})\bigcap(c_n^{min}, c_n^{max})\neq \varnothing$ (see Fig. \ref{fig:dark_render}(c)).

\begin{algorithm}[t]
   \caption{Generating the 2D adversarial texture map}
   
   {\textbf{Input:}}\vspace{6pt}\quad\begin{minipage}[t]{183pt}
     information of the sensor $\Theta=\{r, \theta, \varphi\}$ in every image $\mathbf{I}_{p}$ rendered by PBR, 3D model $\mathbf{M} =(\mathbf{M}_0, \mathbf{T}_0)$, and target class $y$
   \end{minipage}

   {\textbf{Output:}}\quad\begin{minipage}[t]{160pt}
    adversarial 2D texture map $\mathbf{T}_{adv}$
   \end{minipage}
   \label{algorithm}

   \quad\begin{minipage}[t]{220pt}
   \begin{algorithmic} [1]

   \State $\mathbf{T}_{adv} \gets \mathbf{T}_0$
   \For {epochs} 
   \State Match $\Theta$ to images
   \For{num of images}
      \State $\mathbf{M} \gets (\mathbf{M}_0,\mathbf{T}_{adv})$
      \State $\mathbf{I}_{d} \gets \mathcal{R}_d(\Theta, \mathbf{M}, \mathbf{S}')$
      \State Transform textures to get $\mathbf{I}_{d}'$
      \State $\mathbf{I}_{m} \gets \mathcal{R}_d(\Theta, (\mathbf{M}_0, \mathbf{T}_0), \mathbf{S}')$
      \State Calculate $\mathbf{P}$ with $\mathbf{I}_{m}$ by Eqns. (\ref{eq:cmid}) and (\ref{eq:P})
      \State $\mathbf{I}_{o} \gets \mathbf{I}_{d}'*\mathbf{P}+\mathbf{I}_{p}*(1-\mathbf{P})$
      \State Get scores $\mathcal{F}^{cls}, \mathcal{F}^{obj}$ by detecting $\mathbf{I}_{o}$
      \State Calculate $\mathcal{L}_{1}, \mathcal{L}_{2}, \mathcal{L}_{s}$ by Eqns.(\ref{eq:two stage}), (\ref{eq:one stage}), \\ \qquad \qquad\qquad\quad and (\ref{eq:tv})
      \State $\mathcal{L} \gets \mathcal{L}_1+ \beta \mathcal{L}_2+\gamma \mathcal{L}_{s}$
      \State Optimize $\mathbf{T}_{adv}$ by minimizing $\mathcal{L}$
      \EndFor
   \EndFor
   \end{algorithmic}
   
  \end{minipage}
\end{algorithm}

\begin{figure*}[t]
   \begin{center}
   \includegraphics[width=1.0\linewidth]{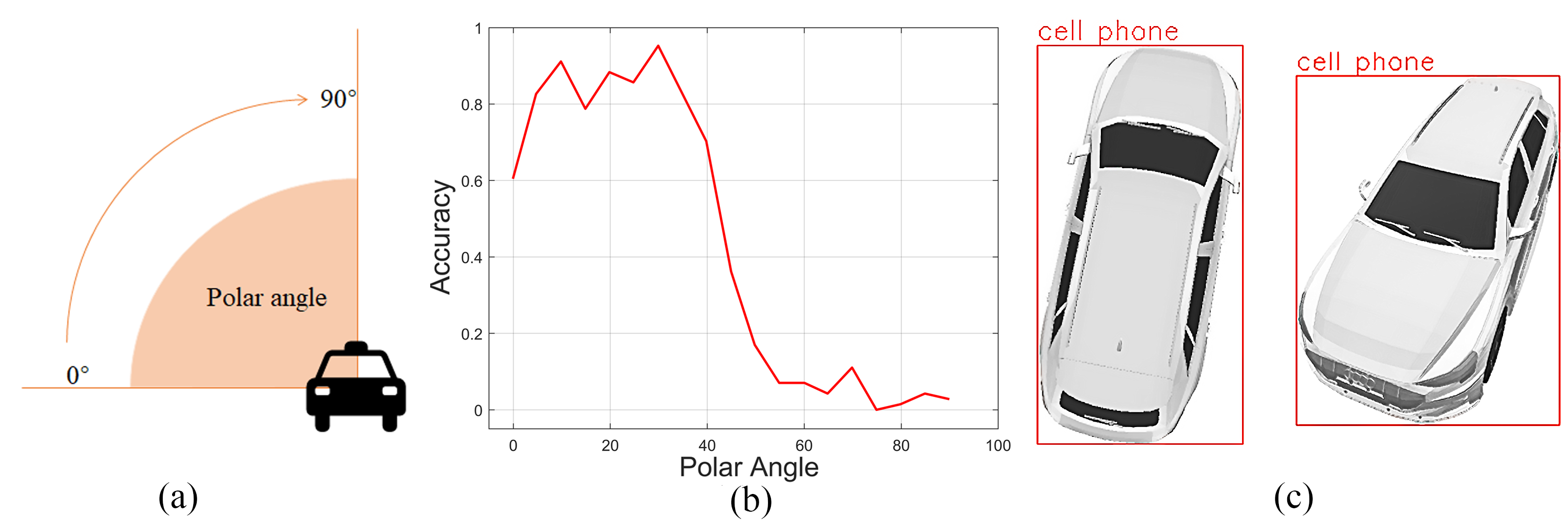}
   \end{center}
      \caption{Pretrained model performance decreases with increasing polar angles. (a) The range of polar angles. (b) The accuracies of the model used by both DAS \cite{wang2021dual} and FCA \cite{wang2022fca} at different polar angles. As shown, the accuracy dropped rapidly when the polar angle increased over $45^{\circ}$. (c) The car was always detected as a "cell phone" at top views (with the conference threshold of 0.9). Here are two examples. }
   \label{fig:drop}
 \end{figure*}

\begin{figure*}[t]
   \begin{center}
   \includegraphics[width=0.95 \linewidth]{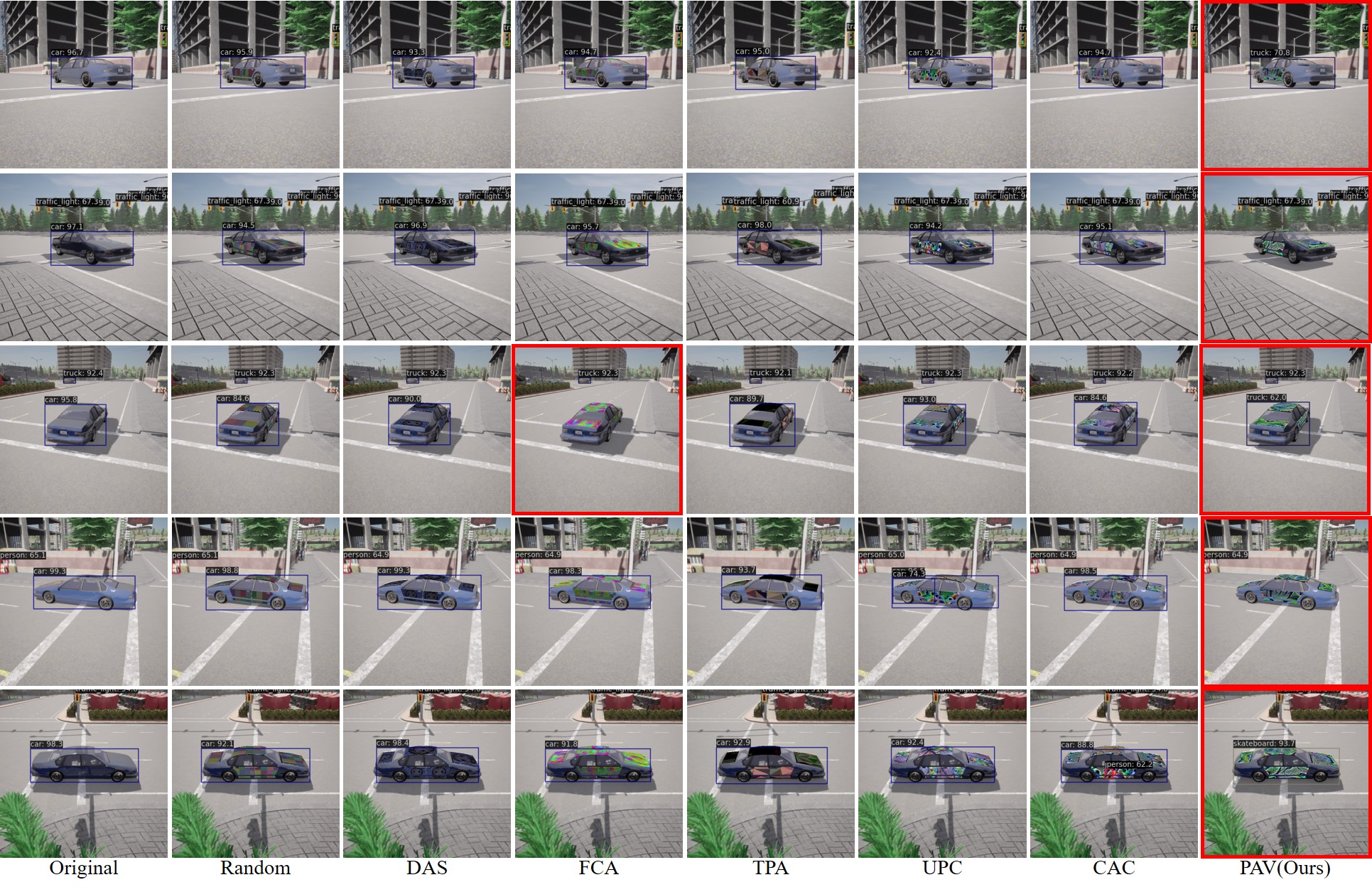}
   \end{center}
   \caption{Some original, random and adversarial examples generated by different methods in the digital world. The images surrounded by red frames mean successfully attacked images. In the digital world, our method has a stronger and more stable attack performance than other methods at various angles.}
   \label{fig:digital}
\end{figure*}

\subsection{Optimization}

Our aim is to make the target vehicle undetected and misclassified by detectors. To achieve it, we attack the classification and detection scores of the target detector simultaneously, so that the classification confidence of the target class (car) and the detection confidence of the target object decrease.

{\textbf{Detector loss.}} The average of the detection confidences $\mathcal{F}_{t}^{obj}(\mathbf{I}_{o})$ is taken as the first part of the attack loss, which is represented as
\begin{equation}
\mathcal{L}_1=\frac{\sum \mathcal{F}_{t}^{obj}(\mathbf{I}_{o})}{n_o},
\label{eq:two stage}
\end{equation}
where $n_o$ denotes the number of bounding boxes. 
The average of the classification confidences $\mathcal{F}_{t}^{cls}$ of all detected boxes with the target class $t$ is used as the second part of the attack loss, 
\begin{equation}
\mathcal{L}_2=\frac{\sum \mathcal{F}_{t}^{cls}(\mathbf{I}_{o})}{n_c},
\label{eq:one stage}
\end{equation}
where $n_c$ denotes the number of proposals.
We minimize the combination of the two functions:
\begin{equation}
   \label{eq:sum12}
\mathcal{L}_a=\mathcal{L}_1+ \beta \mathcal{L}_2,
\end{equation}
where $\beta$ is a hyperparameter.


\textbf{Smoothing loss.}
To generate a natural camouflage instead of a noise-like texture map, we incorporate the smoothing loss \cite{sharif2016accessorize} $\mathcal{L}_{s}$ of the texture map as a component of the optimization objective. The smoothing loss is defined as
\begin{equation}
   \label{eq:tv}
\begin{split}
\mathcal{L}_{s} =\sum((\mathbf{T}_{_{adv}(i,j)}-\mathbf{T}_{_{adv}(i+1,j)})^2+ \\
(\mathbf{T}_{_{adv}(i,j)}-\mathbf{T}_{_{adv}(i,j+1)})^2),
\end{split}
\end{equation}
where $i,j$ denote the $i$-th row, $j$-th column in the texture map $\mathbf{T}_{_{adv}}$.

Then a 3D camouflage with textures is updated by minimizing the loss function $\mathcal{L}$, which is composed of detector loss $\mathcal{L}_a$ and smoothing loss $\mathcal{L}_{s}$:
\begin{equation}
   \label{allloss}
 \mathcal{L}=\mathcal{L}_a+\gamma \mathcal{L}_{s},
\end{equation}
where $\gamma$ is a hyperparameter determined empirically. 

The overall algorithm is summarized in Algorithm \ref{algorithm}.

\section{Experiments}
In this section, we present the experimental results obtained in both the digital and physical worlds.

\subsection{Experimental Settings}

\begin{table*}[t]
   \centering
   \caption{ASR(\%) and P@0.5(\%) of different detectors in the digital world when applying camouflages generated by attacking different detectors. The first row indicates the target detector for white-box attack, and the first column indicates the detector used for testing. Data with * inicates white-box attacks. }
   
   \resizebox{1.0\linewidth}{!}{
   \begin{tabular}{ccccccccccccc}
      \hline 

      \multirow{2}{*}{Detector} & \multicolumn{2}{c}{Faster R-CNN} & \multicolumn{2}{c}{YOLOv3} & \multicolumn{2}{c}{SSD} & \multicolumn{2}{c}{DETR} & \multicolumn{2}{c}{DINO} & \multicolumn{2}{c}{DDQ}\\
      \cmidrule(lr){2-3}\cmidrule(lr){4-5}\cmidrule(lr){6-7}\cmidrule(lr){8-9}\cmidrule(lr){10-11}\cmidrule(lr){12-13}
      & ASR & P@0.5  & ASR & P@0.5  & ASR & P@0.5  & ASR & P@0.5  & ASR & P@0.5  & ASR & P@0.5 \\ 
      \hline

      Faster R-CNN & \textbf{89.8}* & \textbf{9.1}* & 83.1 & 28.0 & 87.2 & 27.6 & 84.1 & 3.2& \textbf{99.2} & 10.0 & \textbf{98.3} & 36.6 \\
      YOLOv3 & 82.6 & 26.4 & \textbf{93.1}* & \textbf{4.0}*  & 86.1 & 28.7 & 87.0 & \textbf{1.7}  & 99.1 & 10.5 & 97.4 & 43.9 \\
      SSD & 72.4 & 35.1 & 89.2 & 16.2 & \textbf{90.4}* & \textbf{12.9}* & \textbf{88.1} & 60.9 & 94.7 & 25.1 & 90.1 & 61.2 \\
      DETR   & 70.1 & 47.9 & 76.2 & 40.5 & 72.7 & 56.2 & 86.3* & 1.9* & 98.6 & 18.2 & 91.3 & 55.2 \\
      DINO   & 76.8 & 29.6 & 81.1 & 32.8 & 80.5 & 40.3 & 80.5 & 39.8 & \textbf{99.2}* & \textbf{6.5}*  & 97.3 & 36.9 \\
      DDQ    & 81.4 & 23.9 & 79.0 & 30.9 & 84.6 & 32.2 & 85.4 & 32.4 & 99.1 & 7.7  & 97.6* & \textbf{20.1}* \\
      \hline
   \end{tabular}
   } 
   \label{tab:transfer}
 \end{table*}

\begin{figure}[t]
  \begin{center}
  \includegraphics[width=1.0\linewidth]{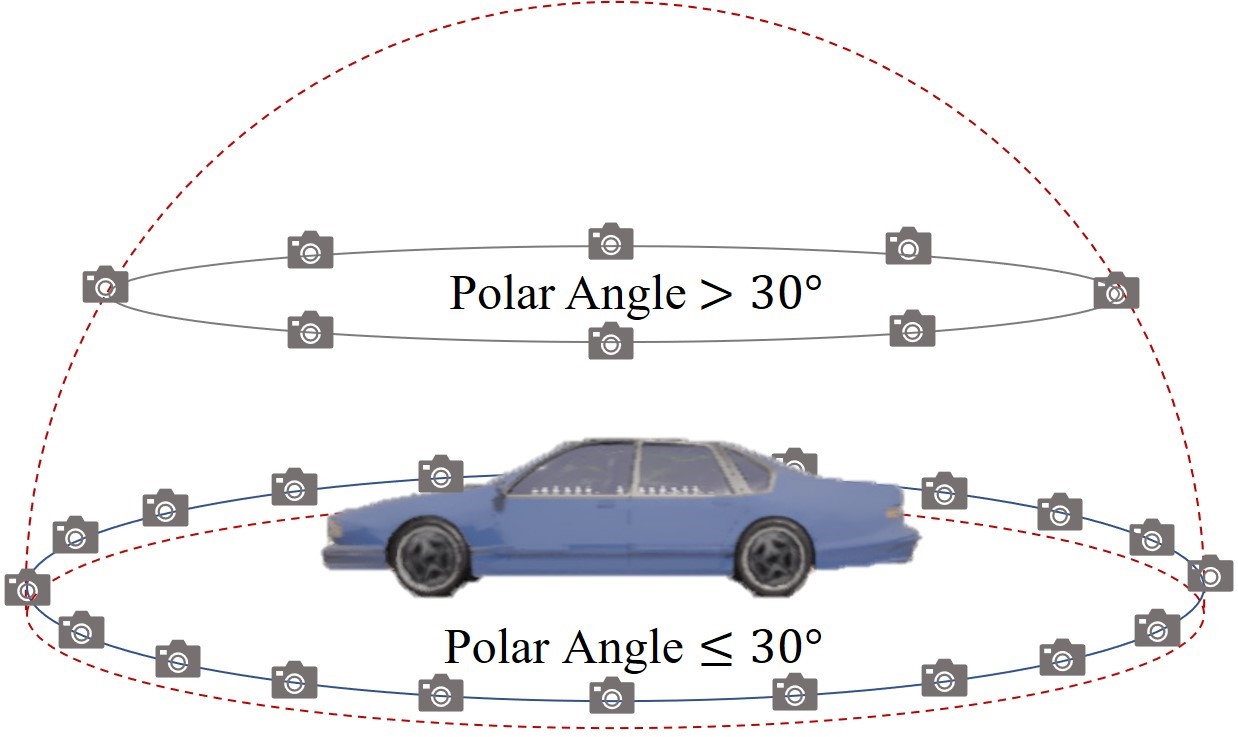}
  \end{center}
     \caption{The distribution of cameras when collecting the dataset in PBR.}
  \label{fig:collect}
  \end{figure}

  \begin{figure}[t]
   \begin{center}
   \includegraphics[width=1.0\linewidth]{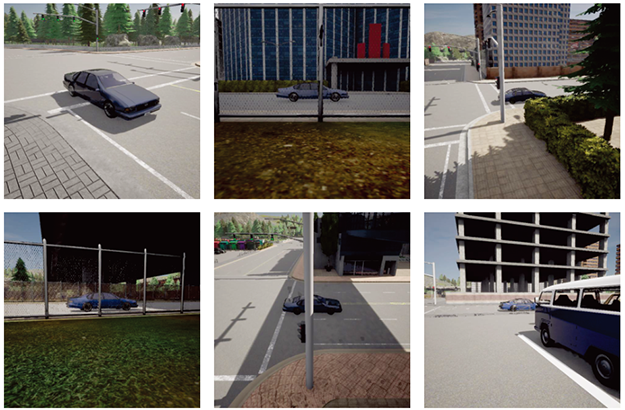}
   \end{center}
   \caption{Sample images in different viewing angles, distances and scenes from the dataset collected. }
   \label{fig:dataset}
\end{figure}

  \begin{figure*}[!h]
   \begin{center}
   \includegraphics[width=0.8\linewidth]{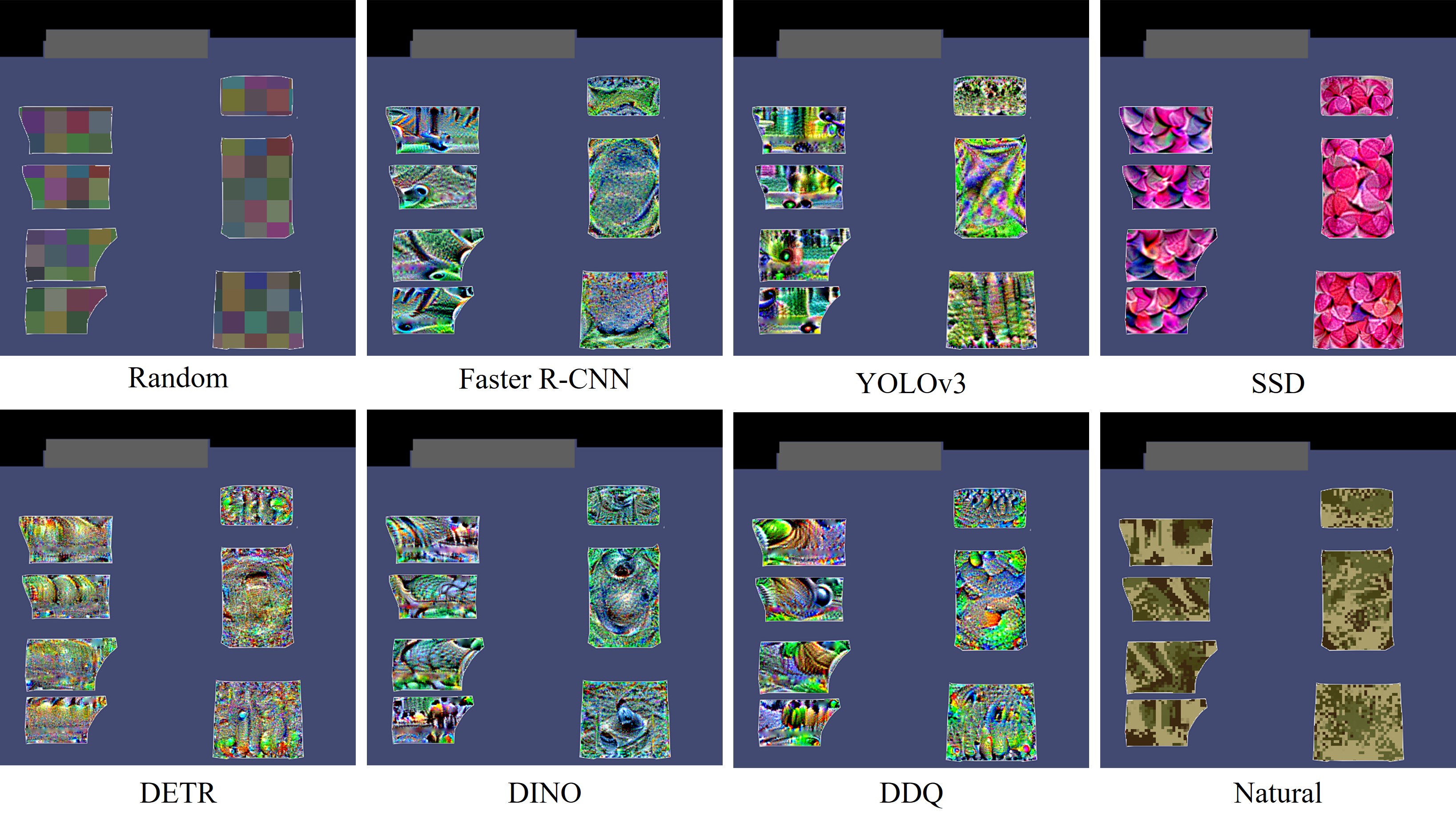}
   \end{center}
   \caption{The random map and PAV adversarial maps under various attack settings. The detector name under each map represents the target model, and 'Natural' represents the natural adversarial map optimized for Faster R-CNN.}
   \label{fig:model_map}
\end{figure*}

\textbf{Datasets.}\quad  Since rendering results of PBR do not need to be involved in the optimizing process, the vehicle could be pre-rendered using PBR to generate the dataset. Although an open dataset created by Wang \textit{et al.} \cite{wang2021dual} contains some rendering results with high polar angles, pretrained DNNs usually inference these data with low accuracies, because top views of vehicles are relatively rare in most existing datasets \cite{lin2014microsoft, deng2009imagenet}. 
The detection results at different polar angles are shown in Fig. \ref{fig:drop}. It is seen that when the polar angle was larger than $45^{\circ}$, the detection accuracy was very low, which is not suitable to demonstrate the attack performance. Therefore, we decided to create a more suitable dataset. 

In this work, we chose CARLA \cite{dosovitskiy2017carla} as a representative for PBR, and 3D Chevrolet Impala vehicle model as the target object. More specifically, within the CARLA world coordinate system, the vehicle was placed in 10 distinct locations within each of the 8 cities, resulting in a total of 80 locations. In the spherical coordinate system, the distance between the camera sensor and the vehicle was set to the values of $(8, 10, 14, 20)$. The values of the camera polar angles were $(5^{\circ},10^{\circ},20^{\circ},30^{\circ},45^{\circ})$. When the polar angle was below $30^{\circ}$, the vehicle features in the camera were sensitive to changes of the azimuthal angle. In other words, even slight changes of the azimuthal angle can result in significant differences in the rendered images. Therefore, the sampling frequency for the azimuthal angle was set relatively high, with an interval of $18^{\circ}$. Conversely, when the polar angle was above $30^{\circ}$, changes of the azimuthal angle had less impact on the rendering results. Therefore, the sampling interval was set to $45^{\circ}$ to avoid data similarity.
We excluded data with polar angles exceeding $45^{\circ}$ to avoid low detection accuracy (shown in Fig. \ref{fig:drop} (b)). The camera placement strategy is illustrated in Fig. \ref{fig:collect}. 

Finally, the images obtained from 72 locations were divided to form the training set, which consisted of a total of 6336 images. The images collected from the remaining 8 locations were then allocated to the testing set containing 704 images. This dataset is named $XFL\_VI\_CITY$\footnote. Sample images in the collected dataset are shown in Fig. \ref{fig:dataset}.

\textbf{Victim detectors.}\quad For our method, we chose Faster R-CNN \cite{ren2015faster}, YOLOv3 \cite{redmon2018yolov3}, SSD \cite{liu2016ssd}, DETR \cite{carion2020end}, DINO \cite{zhang2022dino}, and DDQ \cite{zhang2023dense} as white-box detectors in the experiments. For UPC \cite{huang2020universal}, Faster R-CNN was used as target detector for subsequent fair comparison. 
The methods FCA \cite{wang2022fca} and TPA \cite{zhang2023boosting} were specifically designed for the network architecture of YOLOv3, so we reproduced the method with its original settings. DAS \cite{wang2021dual} was designed to suppress the attention of detectors, so we extracted the attention map of YOLOv3 to attack. For CAC \cite{duan2021learning}, we used the method' s target detector Faster R-CNN. 
These detectors were all pretrained on the dataset COCO \cite{lin2014microsoft}.

\begin{table*}[t]
   \centering
   \caption{ASR(\%) and P@0.5(\%) of different detectors when applying camouflages generated by using different attack methods. Data with * inicates white-box attacks. FCA, TPA are designed for YOLOv3, while UPC, CAC are designed for Faster R-CNN. DAS is based on YOLOv3 features. }
   
   \resizebox{1.0\linewidth}{!}{
   \begin{tabular}{ccccccccccccc}
      \hline
     
      \multirow{2}{*}{Method} & \multicolumn{2}{c}{Faster R-CNN} & \multicolumn{2}{c}{YOLOv3} & \multicolumn{2}{c}{SSD} & \multicolumn{2}{c}{DETR} & \multicolumn{2}{c}{DINO} & \multicolumn{2}{c}{DDQ}\\ 
      \cmidrule(lr){2-3}\cmidrule(lr){4-5}\cmidrule(lr){6-7}\cmidrule(lr){8-9}\cmidrule(lr){10-11}\cmidrule(lr){12-13}
      & ASR & P@0.5 & ASR & P@0.5 & ASR & P@0.5 & ASR & P@0.5 & ASR & P@0.5 & ASR & P@0.5\\ 
      \hline
      Original      & - & 87.4 & - & 83.9 & - & 89.3 & - & 89.3 & - & 87.4 & - & 87.6\\ 
      Random     & 28.0 & 75.4 & 28.5 & 75.7 & 22.3 & 87.3 & 40.4 & 44.1 & 47.1 & 70.6 & 39.6 & 80.2\\ 
      DAS \cite{wang2021dual}   & 28.4 & 80.1 & 39.5* & 75.1* & 38.0 & 75.2 & 30.0 & 62.4 & 70.1 & 67.7 & 65.2 & 75.5\\ 
      FCA \cite{wang2022fca}  & 41.0 & 72.6 & 37.0* & 77.9* & 32.0 & 83.9 & 64.9 & 22.4 & 87.3 & 50.8 & 68.1 & 76.4 \\
      FCA-full & 43.8 & 73.0 & 51.8* & 68.2* & 42.1 & 82.4 & 57.0 & 33.3 & 79.4 & 57.1 & 61.6 & 83.1\\ 
      TPA \cite{zhang2023boosting} & 31.1 & 54.2 & 47.7* & 39.0* & 45.0 & 55.4 & 27.6 & 44.5 & 56.7 & 58.0 & 45.7 & 68.8 \\ 
      UPC \cite{huang2020universal}        & 42.8* & 68.7* & 45.7 & 72.1 & 41.6 & 78.0 & 46.3 & 42.7 & 88.4 & 50.9 & 66.2 & 74.1\\  
      CAC \cite{duan2021learning} & 54.3* & 59.3* & 52.7 & 68.0 & 44.5 & 77.8 & 65.1 & 19.1 & 90.5 & 56.5 & 68.6 & 76.1\\ 
      CAC-full  & 79.1* & 34.1* & 78.4 & 38.1 & 72.4 & 60.7 & 79.9 & 4.3  & 97.3 & 32.4 & 88.9 & 69.4\\ 
      \hline
      PAV(ours)-natural & 76.8* & 36.8* & \textbf{87.3} & \textbf{23.5} & 85.4 & 33.0 & 75.8 & 11.6 & 96.8 & 27.1 & 87.0 & 68.0\\ 
      PAV(ours) & \textbf{89.8}* & \textbf{9.1}* & 83.1 & 28.0 &\textbf{ 87.2} & \textbf{27.6} & \textbf{84.1} & \textbf{3.2} & \textbf{99.2} & \textbf{10.0} & \textbf{98.3} & \textbf{36.6}  \\

      \hline
   \end{tabular}
   }  
   \label{tab:detect}
 \end{table*}

\textbf{Evaluation metrics.}\quad 
In the digital experiments, attack performance was evaluated using common metrics.: Precision of the class "car" at IoU threshold of 0.5 (P@0.5) and Attack Success Rate (ASR). In the physical experiments, recorded videos of the vehicle from different locations and angles were processed frame by frame using the detectors. The number of frames where the vehicle was successfully detected is denoted as $f_d$, and the total number of frames in the video is denoted as $f_o$. The accuracy $A_{physical}$ in the physical world is defined as follows:
\begin{equation}
   \label{eq:asrphy}
   A_{physical} = \frac{f_{d}}{f_{o}}.
\end{equation}

\textbf{Implementation details.}\quad 
Before optimizing the textures, we used MAYA \cite{derakhshani2012introducing} to adjust the coordinates of U-V mapping. In the combination process, we chose PyTorch3D \cite{ravi2020accelerating} as DR. 
To align the rendering images of DR and those of PBR, the sensor coordinates were scaled with the ratio 0.345 and the vehicle coordinate was shifted from $(0, 0, 0)$ to $(0, -0.235, -0.07)$.
The factors $\beta, \gamma$ were set to 1.0 and 0.5 respectively empirically. 
The learning rate of the optimization was set to 0.015 and the number of epochs was 5. We implemented PAV-camou using the Pytorch framework accelerated by NVIDIA RTX 3080 GPU. 
In the physical experiments, our 3D car model was printed using the Stratasys J850, and videos were recorded with Honor V20.


\subsection{Digital World Attack}

In this subsection, we compare the effectiveness of adversarial camouflages in the digital world by using a 1:20 scale 3D model car.

\begin{figure}[h]
   \begin{center}
   \includegraphics[width=0.95 \linewidth]{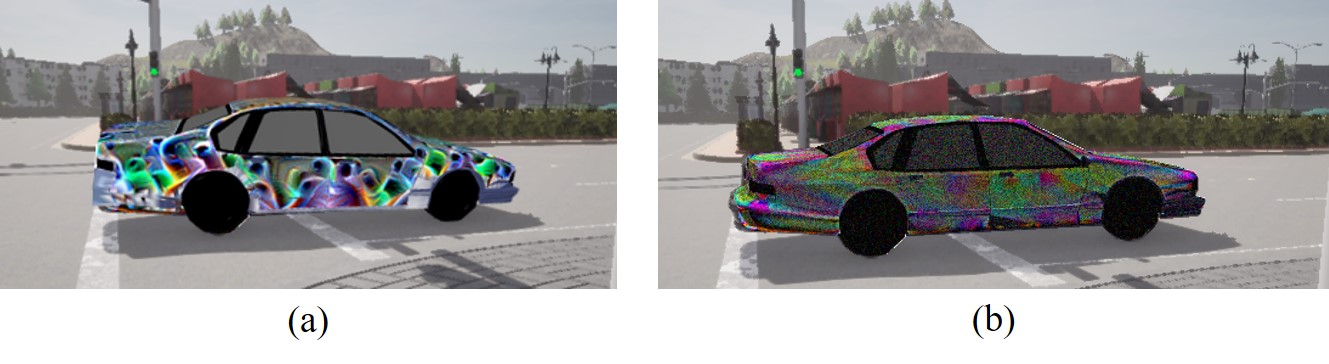}
   \end{center}
   \caption{Full-vehicle camouflage optimized by CAC (a) and FCA (b).}
   \label{fig:full}
\end{figure}
\textbf{Attacking results.} The aforementioned metrics were employed to test the effectiveness of our method on different detectors under the experimental settings described above. The optimized adversarial map under these settings is shown in Fig. \ref{fig:model_map}. See the results in Table \ref{tab:transfer}. 
Each row in the table represents the detection results of the same adversarial texture under different detectors, data marked with an asterisk (*) indicates white-box attack results, while the other data in the same row represent black-box attack results.

In the setting of white-box attack, our method decreased P@0.5 of Faster R-CNN, YOLOv3, SSD, DETR, DINO, and DDQ to 9.1\%, 4.0\%, 12.9\%, 1.9\%, 6.5\%, 20.1\% respectively, and got high ASRs from 86.3\% to 99.2\%. 
These generated camouflages also exhibited good transferability to each other detectors, especially the camouflage targeting Faster R-CNN, which showed the best ASR in black-box attacks against DINO and DDQ. Therefore, in subsequent digital and physical experiments, we conducted analysis using this camouflage. 
Considering that the methods of FCA and CAC aim for full-vehicle camouflage, we also replicated their attack effects under the full-vehicle camouflage setup. The methods marked with '-full' in the table represent full-vehicle camouflage. The optimized camouflage can be seen in Fig. \ref{fig:full}.

\textbf{Comparison with other methods.} We compared the proposed method with several mainstream attack methods that utilized 3D differentiable renderers, namely DAS \cite{wang2021dual}, TPA \cite{zhang2023boosting}, and FCA \cite{wang2022fca}, CAC \cite{duan2021learning}, as well as the UPC \cite{huang2020universal} method based on 2D training. We reproduced DAS and FCA using the 3D Impala vehicle model and applied the optimized textures in the same manner. In order to ensure fairness, the patterns trained by UPC were applied to the same areas of the vehicle in our experiments. We generated a $32\times32$ random image and enlarged it to the texture size by repeating pixels. Then we compared the camouflage optimized by the proposed method with the original textures, random textures, and adversarial camouflages generated by other methods. These maps can be seen in Fig. \ref{fig:model_map}. Furthermore, considering that the attacks in the FCA and CAC were conducted under the setting that the entire car body was camouflaged, we also set our car body to be fully camouflaged and compared it with these two methods. The target detectors and detection results of different attack methods are shown in Table \ref{tab:detect}. 
As shown, our method achieved the best results.

Surprisingly, the methods DAS and TPA, which were also based on 3D training, exhibited poorer attack performance than random textures on certain detectors. 
This is because DAS and TPA do not have 2D mapping textures, so optimization of textures is only possible for the faces in the 3D model, which requires that the number of pixels on each triangular face is the same (the number is 36 in DAS, and 1 in TPA). This causes that larger triangular faces and smaller triangular faces express an equal amount of information, which limits the overall amount of information that can be conveyed in the textured part of the car model. Besides, since most faces in 3D models aren't needed for training, this strategy can can lead to unnecessary memory usage by faces not required for training.
Some examples in the digital world are shown in Fig. \ref{fig:digital}.

\begin{table}[b]
   \caption{ASR(\%) of Faster R-CNN when adopting adversarial camouflage optimized with different components of our loss. Faster means the detector Faster R-CNN.}
   \centering
   \resizebox{1.0\linewidth}{!}{
         \begin{tabular}{ccccccc}
            \hline 
            Loss& Faster & YOLOv3 & SSD & DETR& DINO & DDQ\\
            \hline 
             $\mathcal{L}_1$ & 90.7 & 81.1& 84.9  & 85.8  & 99.1  & 97.3 \\
             $\mathcal{L}_2$  & 89.8 & 80.0& 85.2  & 85.4  & 99.1   & 97.0 \\
             $\mathcal{L}_1+\mathcal{L}_2$ & \textbf{90.7}  & 81.4 & 85.5  & \textbf{87.0} & 99.1   & 97.1 \\
             $\mathcal{L}_1+\mathcal{L}_2+\mathcal{L}_{s}$&  89.8  & \textbf{83.1} &\textbf{ 87.2} & 84.1  & \textbf{99.2}  & \textbf{98.3} \\ 

            \hline
         \end{tabular}
   }
   \label{tab:our_loss}
\end{table}

\textbf{Effectiveness of the proposed loss function.} We investigated the effectiveness of each component in our loss function separately (see Eqn. (\ref{eq:sum12}) and (\ref{allloss})). The results are shown in Table \ref{tab:our_loss}. Using either $\mathcal{L}_1$ or $\mathcal{L}_2$, ASR was at least 89.8\% on Faster R-CNN. 
Though both $\mathcal{L}_1$ and $\mathcal{L}_2$ contributed to reducing the accuracy of the model, attacking the detection box confidence $\mathcal{L}_1$ yielded better results compared to the classification confidence $\mathcal{L}_2$. A slight improvement was also observed in ASR when attacking both $\mathcal{L}_1$ and $\mathcal{L}_2$. 

\begin{table}[ht!]
   \caption{ASR(\%) of Faster R-CNN before and after adding adversarial defense methods.}
   \centering
      \begin{tabular}{cc}
         \hline 
         Method  & ASR(\%)   \\
         \hline 
         w/o defense methods& 89.8 \\
         Adversarial Training& 75.6\\
         Spatial Smoothing& 85.2\\ 
         Feature Norm Clipping& 81.4\\
         Feature Squeezing& 87.0\\ 
         Cut Out          & 89.9\\ 
         \hline
      \end{tabular}
   \label{tab:defense}
\end{table}

\begin{figure*}[t]
   \begin{center}
   \includegraphics[width=1.0\linewidth]{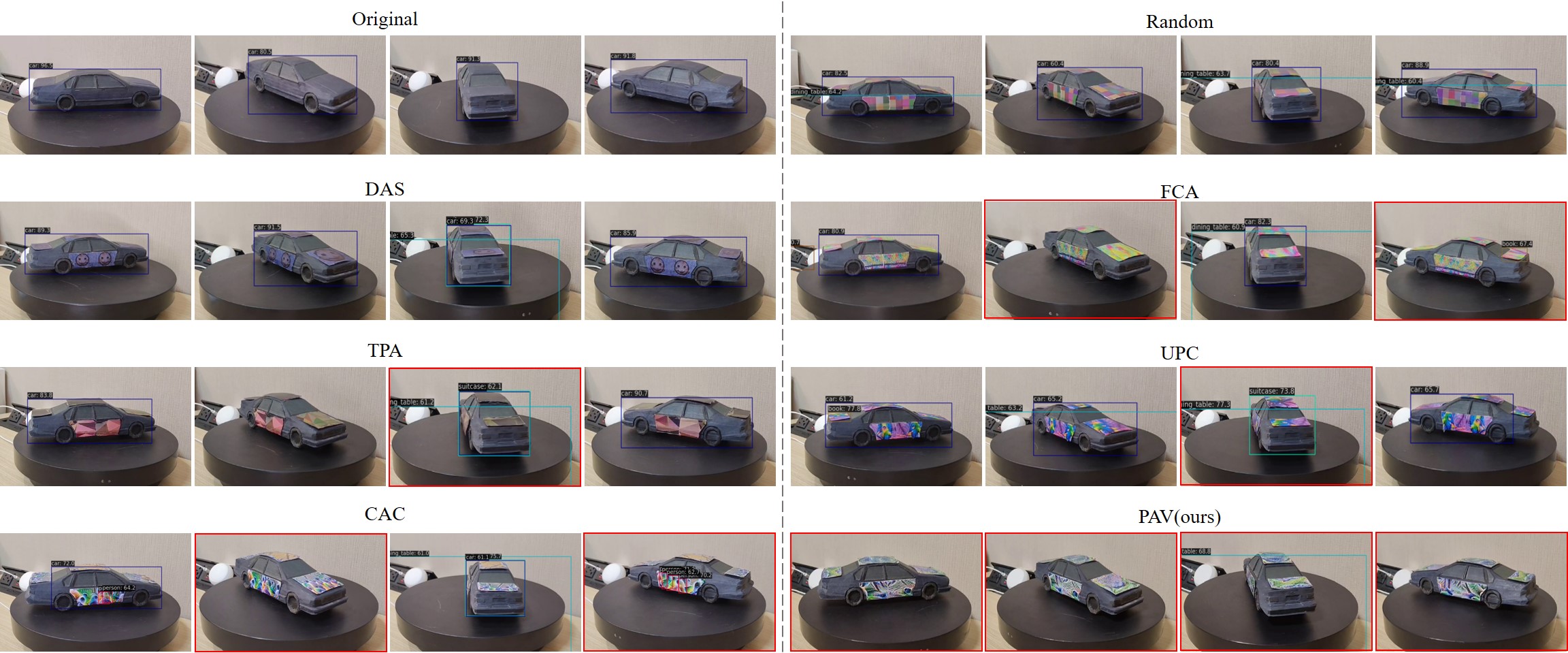}
   \end{center}
      \caption{The physical experimental examples of different attack methods. The text above each row of images represents the name of the adversarial attack method. In the physical world, DAS and FCA showed poor performance in the physical world, UPC did not fool detectors in multiple viewing angles, while our method kept the effectiveness and hided the car at most viewing angles in the physical world. The threshold of the detectors was set to 0.5. The images enclosed in red boxes indicate detection failures.}
   \label{fig:physical}
   \end{figure*}

\begin{table*}[t]
   \caption{The accuracies $A_{physical}$ of different attack methods in the physical world at the polar angles of $0^{\circ}$ and $20^{\circ}$. }
   \centering
   \resizebox{0.65\linewidth}{!}{
   \begin{tabular}{ccccccccccccc}
      \hline 
 
      Method& Faster R-CNN & YOLOv3 & SSD & DETR & DINO &DDQ \\
      \hline
      Original         & 81.8 & 62.5 & 84.1 & 80.7 & 58.0 & 78.4 \\ 
      Random        & 81.8 & 44.3 & 95.5 & 69.3 & 45.5 & 70.5 \\ 

      DAS \cite{wang2021dual} & 77.3 & 51.1 & 87.5 & 58.0 & 28.4 & 60.2 \\ 
      FCA \cite{wang2022fca}& 68.2 & 28.4 & 87.5 & 54.5 & 5.7  & 69.3 \\ 
      TPA \cite{zhang2023boosting}& 73.9 & 34.1 & 85.2 & 46.6 & 45.5 & 64.8 \\ 
      UPC \cite{huang2020universal}& 70.5 & 67.0 & 80.7 & 50.0 & 42.0 & 86.4 \\ 
      CAC \cite{duan2021learning}& 61.4 & 4.5  & 85.2 & 34.1 & 31.8 & 39.8 \\ 
      \hline 
      PAV(ours)-natural & 64.8 & 28.4 & 84.1 & 54.5 & 45.5 & 53.4 \\ 
      PAV(ours) & \textbf{9.1}  & \textbf{1.1}  & \textbf{70.5} & \textbf{20.5} & \textbf{12.5} & \textbf{14.8 }\\ 
      
      \hline

      \hline
   \end{tabular}
   }
   \label{tab:phy}
 \end{table*}

\textbf{Evaluation of defense methods.} We evaluated the effectiveness of our camouflage method on Faster R-CNN equipped with some defense methods. The tested defense methods included Adversarial Training \cite{goodfellow2014explaining}, Spatial Smoothing \cite{xu2017feature}, Feature Norm Clipping \cite{yu2021defending}, Feature Squeezing \cite{xu2017feature} and Cut Out \cite{devries2017improved}. 
The results are presented in Table \ref{tab:defense}. Among the defense methods, Adversarial Training performed better than others, but the lowest ASR across the results was 75.6\%. This suggests that the proposed method has a strong attack effect on detectors with defense processing.



\subsection{Physical World Attack}

We evaluated the attack performance of the textures in the physical world by using a 1:20 scale model car (Chevrolet Impala SS 1996).

%
\textbf{Physical experiment settings.} We realized the optimized adversarial camouflages of each method. For our method, we printed the 2D adversarial map generated by attacking Faster R-CNN, which performed the best in the digital world (see Table \ref{tab:detect}). Then we tested the black-box attack effectiveness on other detectors. 
For the other 3D adversarial attack methods, we employed the camouflages in the setup of Table \ref{tab:detect} for physical experiments. Due to the independent textures on each triangular mesh face, they could only be realized by printing rendered images. Then the textured parts on the vehicle were clipped from the images. UPC expressed textures by 2D patches, so we printed them directly. Considering fairness, in our physical experiments, we print textures optimized by all methods with the same size and shape.

After printing, we stuck these textures to the car model placed on a rotating platform and recorded videos with a fixed camera. The platform rotated $360^{\circ}$ in approximately 28 seconds, and the video was captured at a frame rate of 30 frames per second. The video frames were then fed into the detector, and the accuracy $ A_{physical}$ of detection in the physical world was calculated by Eqn. (\ref{eq:asrphy}). The results of sampled frames are shown in Fig. \ref{fig:physical}. 

%
\textbf{Comparisons with other methods.} Our method shows the highest attack performance among all methods. For example, it reduces the accuracies of Faster R-CNN to 9.1\% and YOLOv3 to 1.1\%, etc., which are the lowest accuracies compared with other methods. See results in Table \ref{tab:phy}. Some adversarial examples are presented in Fig. \ref{fig:physical}. 

\subsection{Ablation Studies}
To verify the effectiveness of the two modules in our approach, we conducted ablation studies in the physical world (our two modules were mainly designed for physical experiments, and it was difficult to evaluate their performance in the digital world, so we only conducted ablation studies in the physical world).

\begin{figure}[h]
   \begin{center}
   \includegraphics[width=1.0\linewidth]{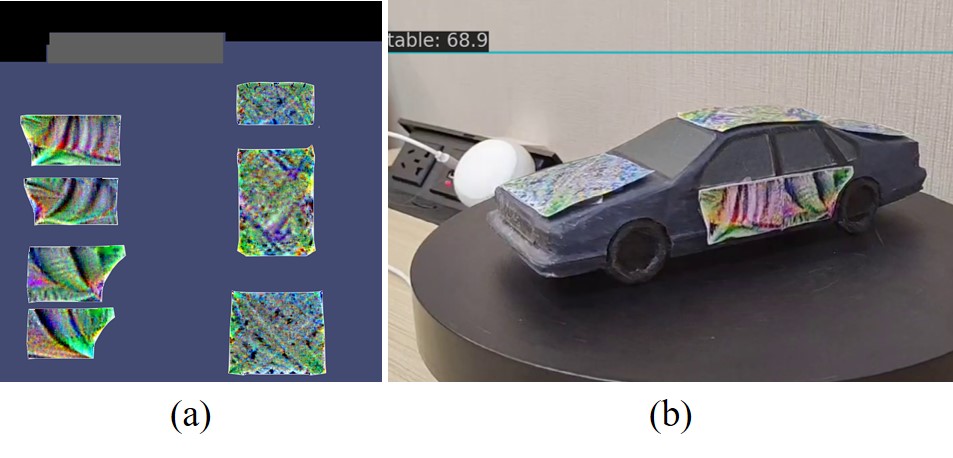}
   \end{center}
      \caption{
         The adversarial camouflage optimized with DR only (a) and an adversarial example in the physical world (b).
      }
      \setlength{\belowcaptionskip}{0 pt}
      \setlength{\abovecaptionskip}{-10 pt}
   \vspace{-5 pt}
   \label{fig:tex}
   \end{figure}

\textbf{Results with DR only.} We aimed to compare the effects of two rendering strategies on attacking detectors in the physical world. Since the optimization required DR, we retained DR for the rendering process and removed PBR to evaluate the impact of its absence on attack effectiveness.
Differences in the rendering strategies used led to differences in the optimized textures. Fig. \ref{fig:tex} shows the adversarial map optimized with DR only and an adversarial example in the physical world. Compared with the texture optimized by the dual-renderer combination strategy (see Fig. \ref{fig:model_map}), the adversarial texture optimized under this setting looks more "blurry". We also conducted a quantitative comparison to evaluate two camouflages in the physical world. The experimental results are presented in Table \ref{tab:ablation}. It is observed that the camouflage optimized using the two-renderer combination strategy exhibited better attack effectiveness.

\begin{figure}[h]
   \begin{center}
   \includegraphics[width=0.9\linewidth]{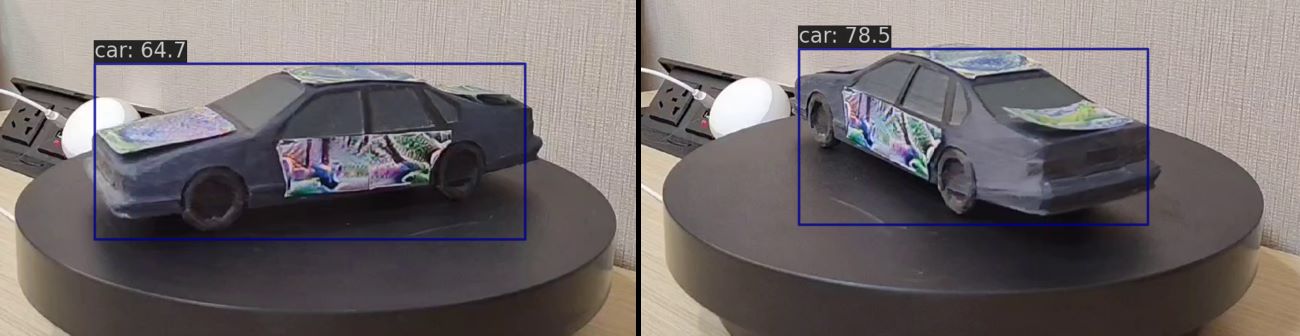}
   \end{center}
      \caption{
         Some examples of the camouflage optimized without reasonable U-V coordinate adjustment. 
      }
      \setlength{\belowcaptionskip}{0 pt}
      \setlength{\abovecaptionskip}{-10 pt}
   \vspace{-5 pt}
   \label{fig:trans}
   \end{figure}

\textbf{Effect of coordinate adjustments.} 
Discarding the U-V coordinate adjustment causes some subtle distortions in the physical printed map. Although these distortions are not easy to detect by the human eye, they lead to reduced attack performance. See some examples in Fig. \ref{fig:trans}. The accuracy drop is also lower than that of camouflage with U-V adjustment, which can be seen in Table \ref{tab:ablation}.

\begin{table}[h]
   \setlength{\abovecaptionskip}{0cm}  
   \caption{The accuracies $A_{physical}$ of methods when adopting camouflages optimized with two modules (PAV), only two-renderer combination module (PAV(W/O U-V)) or and only U-V coordinate adjustment module (PAV(W/O dual)) in the physical world. }
   \centering
   
   \resizebox{1.0\linewidth}{!}{
      \begin{tabular}{ccccccc}
         \hline 
         Method    & Faster R-CNN &  YOLOv3  & SSD   & DETR & DINO &DDQ \\
         \hline 
         PAV      & \textbf{9.1}   & \textbf{1.1}  & \textbf{70.5} & \textbf{20.5} & \textbf{12.5} & \textbf{14.8 } \\ 
         PAV(W/O Com)   & 50.0 & 78.4 & 77.3 & 39.8 & 31.8 & 28.4 \\ 
         PAV(W/O U-V)    & 45.5 & 79.5 & 81.8 & 47.7 & 31.8 & 23.9 \\ 
         \hline
      \end{tabular}
   }
   \label{tab:ablation}
\end{table}

\subsection{Naturalness Studies}
\begin{figure}[h]
   \begin{center}
   \includegraphics[width=0.9\linewidth]{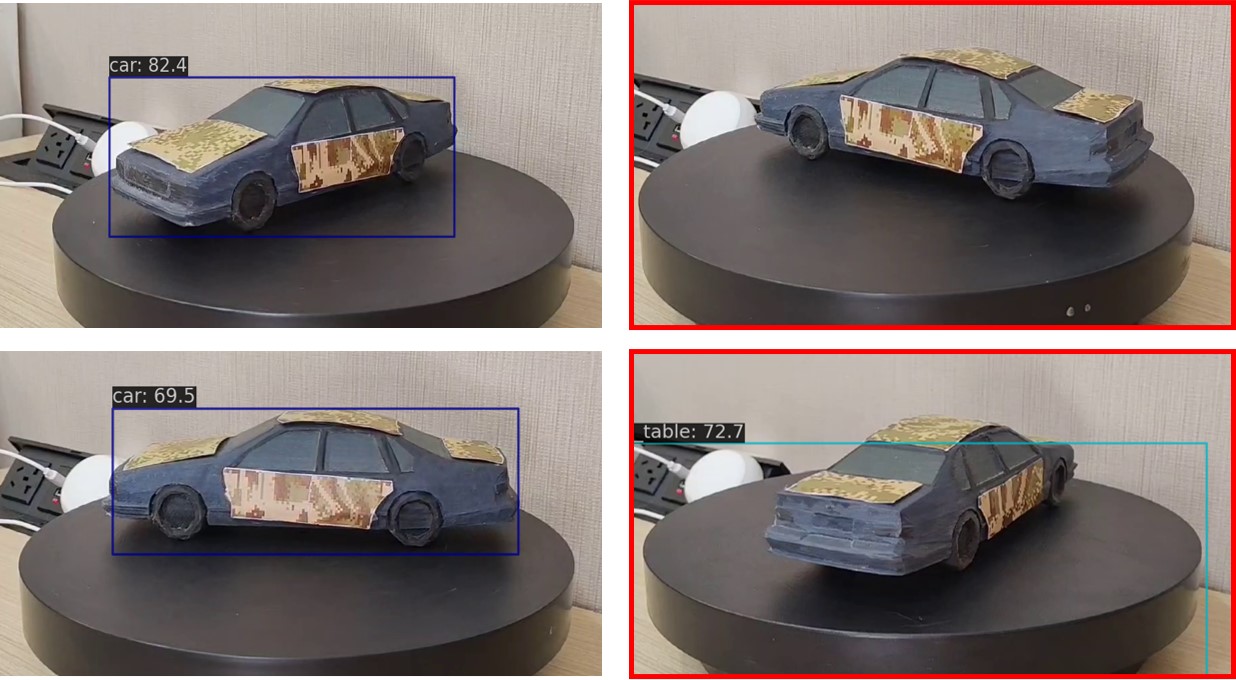}
   \end{center}
      \caption{Some examples of our natural camouflage in the physical world. The left side shows examples of failed attacks and the right side shows successful ones.}
      \setlength{\belowcaptionskip}{0 pt}
      \setlength{\abovecaptionskip}{-10 pt}
   \vspace{-5 pt}
   \label{fig:natural}
   \end{figure}

To improve the naturalness of the textures, we adopted the topology optimization strategy proposed by Hu \textit{et al.} \cite{hu2023physically}, and tested the attack performance. We refer to the camouflage optimized under this setup as PAV-natural, which can be seen in Fig. \ref{fig:model_map}. The experimental results of this camouflage in the digital world and the physical world are presented in Table \ref{tab:transfer} and Table \ref{tab:phy}, respectively.
even with the addition of naturalness constraints, our method still maintains its effectiveness. Nevertheless, it loses some adversarial performance under some specific viewing angles. Some examples are shown in Fig. \ref{fig:natural}.

\section{Conclusion}
We proposed a camouflage-based attack method, which generated a realizable and robust camouflage. Unlike existing methods that render textures onto vehicles directly, we constructed a more reasonable coordinate mapping to optimize 2D textures, resulting in minimal distortion after mapping.
Furthermore, we combined the rendering results of PBR and DR. The non-textured areas on the vehicle were rendered in a more realistic manner, which helped the generated adversarial textures maintain better robustness in the real world.
Compared to previous attack methods, the camouflage of our method achieved higher attack performance in both the digital and physical worlds. Even with the increased naturalness limitation, our method still demonstrated good attack performance.

\section*{Acknowledgment}
This work was supported by the National Natural Science Foundation of China under Grant U2341228.

\bibliographystyle{IEEEtran}
\bibliography{main}

\end{document}